\documentclass[nofootinbib,aps,prd,twocolumn,superscriptaddress,groupedaddress]{revtex4-1}
\tighten
\input epsf
\usepackage{graphicx}% Include figure files
\newcommand{\beq}{\begin{equation}}
\newcommand{\beqa}{\begin{eqnarray}}
		  \newcommand{\eeq}{\end{equation}}
\newcommand{\eeqa}{\end{eqnarray}}

\newcommand{\lsim}{\lesssim}

\newcommand{\lmk}{\left(}
\newcommand{\rmk}{\right)}

\newcommand{\p}{\partial}

\begin{document}

%%%%%%%%%%%%%%%%%%%%%%%%%%%%%%%%%%%%%%%%%%%%%%%%%%%%%%%%%%%%%%%%%%%%%%%%%%
%%%%%%%%%%%%%%%%%%%%%%%%%%%%%%%%%%%%%%%%%%%%%%%%%%%%%%%%%%%%%%%%%%%%%%%%%%
\title{Eccentricity Boost of Stars Around Shrinking Massive Black Hole Binaries  } 
%%%%%%%%%%%%%%%%%%%%%%%%%%%%%%%%%%%%%%%%%%%%%%%%%%%%%%%%%%%%%%%%%%%%%%%%%%
%%%%%%%%%%%%%%%%%%%%%%%%%%%%%%%%%%%%%%%%%%%%%%%%%%%%%%%%%%%%%%%%%%%%%%%%%%
%
%
%%%%%%%%%%%%%%%%%%%%%%%%%%%%%%%%%%%%%%%%%%%%%%%%%%%%%%%%%%%%%%%%%%%%%%%%%%
\author{Mao Iwasa and Naoki Seto}
%%%%%%%%%%%%%%%%%%%%%%%%%%%%%%%%%%%%%%%%%%%%%%%%%%%%%%%%%%%%%%%%%%%%%%%%%%
\affiliation{Department of Physics, Kyoto University
Kyoto 606-8502, Japan
}
%%%%%%%%%%%%%%%%%%%%%%%%%%%%%%%%%%%%%%%%%%%%%%%%%%%%%%%%%%%%%%%%%%%%%%%%%%
\date{\today}

%\baselineskip 11.1mm
%95.85.Sz, 95.30.Sf 
%
%
%
%
%%%%%%%%%%%%%%%%%%%%%%%%%%%%%%%%%%%%%%%%%%%%%%%%%%%%%%%%%%%%%%%%%%%%%%%%%%\begin{abstract}
\begin{abstract}

Based on a simple geometrical approach, we analyze the evolution of the Kozai-Lidov mechanism for stars around shrinking massive black hole binaries on circular orbits. 
We find that, due to a peculiar bifurcation pattern induced by the Newtonian potential of stellar clusters,  the orbit of stars  could become highly eccentric. This transition occurs abruptly for stars with small initial eccentricities. Our approach would be also  useful for studying the Kozai-Lidov mechanism in various astrophysical contexts.

\end{abstract}
\maketitle

\section{introduction}

Coalescence of a massive black hole (MBH) binary is a spectacular event and a huge amount of energy is converted into gravitational waves (GWs). Depending on the masses of the binary, the planned space interferometers such as eLISA \cite{AmaroSeoane:2012km} can detect the strong GW signal, even from a very high redshift. Meanwhile, associated with shrinkage of MBH binaries in galactic nuclei, the infall rate of stars (more precisely, stellar-mass objects) to the MBHs could be enhanced significantly \cite{Chen et al.(2011)}. These stars would emit GWs or electromagnetic waves that encode precious information for astrophysics, cosmology and  fundamental physics ({\it e.g.} strong gravity) \cite{Merritt(2013)}.

For the infall into the MBHs, the star must have an eccentricity close to unity, and it has been actively discussed that the Kozai-Lidov (KL) mechanism could play important roles \cite{Ivanov et al.(2005),Bode & Wegg(2014),Li et al.(2015)}. In the KL mechanism, the inner eccentricity of a hierarchical triple system  increases due to the exchange of inner and outer angular momenta  \cite{Kozai:1962zz}. 

  In this paper, concentrating on the MBH inspirals,  we theoretically  examine the time evolution of the hierarchical triple systems that specifically satisfy the following two conditions; (1) the KL mechanism is initially suppressed by the inner apsidal precessions (due to non-Keplerian potentials \cite{Holman et al.(1997),Ford et al.(2000),Naoz et al.(2013)}), but (2) it gradually becomes effective, along with the slow 
%%@
(to be more precise, adiabatic) contraction of the outer orbits (see also \cite{Shappee & Thompson(2013),Michaely & Perets(2014)}).

Our primary aim is to develop a simple geometrical approach for the adiabatic evolution of the KL mechanism 
%associated with inspirals of  MBH binaries 
(see also \cite{mmr} for  similar studies mean-motion resonances).  More specifically, we analyze the phase-space structure of the corresponding time-dependent Hamiltonian, and concurrently apply the arguments of the adiabatic invariance.  This approach would be complementary to numerical studies, for understanding the KL mechanism of the triple systems including MBH binaries.
 The KL mechanism has been  also applied to various astrophysical phenomena  \cite{Holman et al.(1997),Wen(2003),Ford et al.(2000),Katz & Dong(2012),Fabrycky & Tremaine(2007),Naoz & Fabrycky(2014),Blaes et al.(2002),Naoz et al.(2013),Thompson(2011),Munoz 
& Lai(2015)}, and 
our approach could be useful for studying them. 

This paper is organized as follows.  In section II,   we analyze the hierarchical systems composed purely by three point-masses, nevertheless including relativistic effects.  These simple systems are suitable not only for  explaining our geometrical approach but also for numerically examining  its validity.  Then, in section III, we apply our approach to individual stars in a stellar cluster associated with slowly shrinking MBH binaries. We find that the eccentricities of the stars show intriguing evolutions. These results could be well explained by a peculiar bifurcation pattern caused by the Newtonian potential of the stellar cluster.  We also discuss the effects of relaxation processes for our geometrical approach.
%In this work, we first develop a simple geometrical approach for the KL mechanism in relativistic triple systems, and clarify the impacts of the varying phase space structure (see also \cite{mmr} for mean-motion resonances). Next, we show an  astrophysically important yet overlooked process for individual stars in a stellar cluster associated with slowly shrinking MBH binaries. The eccentricity of the star can generally reach the maximum value allowed by an integral of motion. 
%Furthermore, this transition occurs abruptly for stars with small initial eccentricities. 

\section{KL mechanism for PN systems}
\subsection{Numerical Experiments}

We first examine the evolution of relativistic hierarchical triple systems, only including gravitational interaction with the Post-Newtonian (PN) approximation.  
This is for demonstrating our geometrical approach and also providing an example in distinction from the Newtonian effects of a stellar cluster (discussed in section III).

Our triple systems are made from two spinless MBHs (masses: $m_{0}$, $m_{2}$) and a stellar mass object ($m_{1}$). The inner binary is composed by $m_{0}$ and $m_{1}$, while the remaining MBH $m_{2}$ is the outer tertiary. 
Below, we use the geometrical unit with $c=G=m_{0}+m_{1}+m_{2}=1$ and the labels $j\in\{1,2\}$ for the inner and outer orbital elements. We denote the semimajor axes by $a_{j}$, the eccentricities by $e_{j}$, the arguments of pericenters by $g_{j}$ and the inclination between the inner and outer orbits by $I$. For the inner angular momentum, we also define 
\beq
G_1\equiv \sqrt{1-e_1^2}
\eeq and 
\beq
J_1\equiv G_1\cos I
\eeq
 (the component normal to the outer orbital plane, $J_{1}\leq G_{1}\leq1$).
 To simplify our arguments, we fix  $m_{0}=0.3$, $m_{1}=0.3\times10^{-6}$ and $m_{2}=0.7$ 
%%@
%\footnote{Our unit length and unit time  correspond to $5\times10^{6}(m_{1}/1M_{\odot})\rm km$
% and 17$(m_{1}/1M_{\odot})\rm sec$ in the physical units (scaled with $m_{1}$).}.

To begin with, we make numerical simulations for the relativistic triple systems.
For the direct three-body calculations, we use the equations of motion derived from the 2.5PN ADM Hamiltonian  \cite{Jaranowski:1996nv}, formally expressed as 
\beq
{H}_{\rm ADM}={H}_{\rm N}+{H}_{\rm 1PN}+{H}_{\rm 2PN}+Y{H}_{\rm 2.5PN}
\eeq Here ${H}_{\rm N}, {H}_{\rm 1PN}$ and ${H}_{\rm 2PN}$ are the Newtonian,  1PN, and  2PN terms. The last one ${H}_{\rm 2.5PN}$ is the leading dissipative term and $Y$ is an artificial parameter to speed up our numerical calculations.

Using  a fourth-order Runge-Kutta scheme with $Y=10^4$  \footnote{
We obtained almost the same results with $Y=500$, by dropping  the time-consuming  2PN term.}, we numerically solve the evolution of the following three systems, s-I, s-I\hspace{-.1em}I and s-I\hspace{-.1em}I\hspace{-.1em}I. Their initial conditions are $(e_{\rm 1,i}, a_{2,\rm i},e_{\rm 2,i})=(0.001,6300,0.0001)$ for s-I, (0.001,6300,0.3) for  s-I\hspace{-.1em}I  and (0.3,5700,0.0001) for s-I\hspace{-.1em}I\hspace{-.1em}I. Other orbital elements are commonly set at  $a_{1, \rm i}=400$,  $g_{1,\rm i}=\pi/2$ and  $g_{2, \rm i}=0$. We also choose the initial inclination $I_{\rm i}$ to realize $J_{\rm 1,i}\equiv\sqrt{1-e_{1,\rm i}^{2}}\cos I_{\rm i}=0.150$ ($I_{\rm i}\sim80^\circ$ for our systems).  Our main targets are s-I and s-I\hspace{-.1em}I\hspace{-.1em}I with  different initial inner eccentricities $e_{1,\rm i}$. The system s-I\hspace{-.1em}I is for briefly studying effects of the outer eccentricities $e_{2,\rm i}$.  Note also that we set the above orbital parameters for demonstration of our geometrical approach. More realistic systems  (but more complicated for direct calculations)  would be analyzed in the next section. 

We terminate our calculations, when the instantaneous separation between $m_{0}$-$m_{1}$ becomes smaller than 10$m_{0}$.  All the three systems  are dynamically stable until this termination condition (satisfying the stability criterion of \cite{Mardling & Aarseth(2001)}).

Here, we summarize the characteristic timescales. 
 For the outer orbits of s-I and s-I\hspace{-.1em}I\hspace{-.1em}I, we have the merger time  $T_{2, \rm GW}\simeq4.82\times10^{10}(a_{2}/6000)^{4}(10^{4}/Y)$ due to GW emission \cite{Peters:1964zz}. In the following, we use the decreasing distance $a_{2}$ as an effective time variable. Meanwhile, in the range $a_{2}\geq4000$ relevant for the analysis below, we can neglect the radiation reaction for the inner orbits with $T_{1,\rm GW}>60T_{2,\rm GW}$. The timescales of the KL mechanism and the PN apsidal precession are also important for the inner binary. Their typical values are 
\beqa
T_{\rm KL}&\equiv&\frac{2}{3\pi}\frac{(m_{0}+m_{1}+m_{2})}{m_2}\frac{P_{2}^{2}}{P_{1}},\\
T_{\rm1PN}&\equiv& \frac{a_{1}^{5/2}(1-e_{1}^{2})}{3(m_{0}+m_{1})^{3/2}},
\eeqa
where $P_{1}$ and $P_{2}$ are the inner and outer orbital periods \cite{Holman et al.(1997)}. As we have $T_{2,\rm GW}\gg T_{\rm KL}, T_{\rm 1PN}$ for our three  systems,  the outer orbits shrink adiabatically. Also, the KL mechanism is initially suppressed by the inner relativistic precession with $T_{\rm KL}>T_{\rm 1PN}$.

\begin{figure}[t]
\begin{center}
\includegraphics[width=8.4cm]{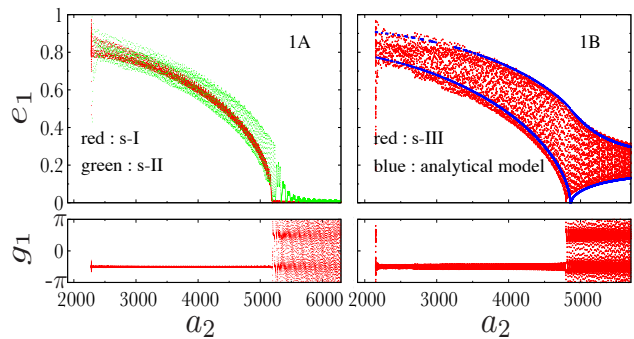}
\end{center}
\vspace*{-0.2cm}
\caption{
Evolution of the inner orbital elements $e_{1}$ and  $g_{1}$, as a function of  decaying $a_{2}$. Panel 1A: results from the direct three-body calculations for s-I (red points) and s-I\hspace{-.1em}I (green points). Panel 1B: results from the direct three-body calculation  for  s-I\hspace{-.1em}I\hspace{-.1em}I (red points) and the corresponding analytical predictions (blue points) for the minimum and maximum of $e_1$   at each $a_{2}$. }
\label{fig:orbit elements}
\end{figure}

Next, we describe our numerical results for s-I and  s-I\hspace{-.1em}I\hspace{-.1em}I. As expected, the shrinkages of $a_1$ are small with $|\delta a_1/a_{\rm 1,i}|<0.04$ down to the epoch of $a_2=4000$ (fluctuations of $J_1$: $|\delta J_1/J_{\rm 1,i}|<0.1$).
Figure. 1 represents the evolution of $e_{1}$ and $g_{1}$ as a function of $a_{2}$. In the case of s-I, $e_{1}$ suddenly starts to increase at $a_{2}\sim5200$ (showing the characteristic feature of a pitchfork bifurcation \cite{strogatz}), and, concurrently,  $g_{1}$ turns  from circulation to libration. 

For s-I\hspace{-.1em}I\hspace{-.1em}I, $e_{1}$ initially oscillates, but it momently vanishes at $a_{2}\sim4800$. Subsequently, $e_{1}$ starts to increase, and $g_{1}$ settles  into libration. 
In Fig. 1A,  the evolution of the system  s-I\hspace{-.1em}I is similar to s-I.  Therefore, the basic aspects of the transition could be understood by analyzing the simpler system s-I with $e_2=0$ (as we limit below).
In common with the three systems,  the mean values of $e_{1}$ evolve slowly with time.

It has been known that, for $e_2\ne0$, the inner eccentricity $e_1$ can experience a resonant-like
 excitation around the epoch with $T_{\rm KL}\sim T_{\rm 1PN}$ \cite{Ford et al.(2000),Naoz et al.(2013)}. Recently, Liu, Lai and Yuan \cite{2015PhRvD..92l4048L} reported that this excitation is caused by an apsidal precession resonance  ${\dot \varpi}_1- {\dot \varpi}_2\simeq 0$. 
Here, $\varpi_i(\equiv \Omega_i+g_i)$ is the longitude of the pericenter and $\Omega_i$ is the longitude
 of the ascending node. 

Our primary target in this paper is the inner orbital evolution for $e_2=0$.  But, here, we briefly comment on this resonance. 
For the run  s-I\hspace{-.1em}I mentioned above, we have ${\dot \varpi}_1\gg {\dot \varpi}_2$ due to the hierarchies of the masses and orbits. Therefore, the resonant state ${\dot \varpi}_1- {\dot \varpi}_2\simeq 0$ was not realized in the run. 

In the next section, we study a more realistic system including a stellar cluster potential with $e_2=0$. It would be interesting to examine the effects of the apsidal precession resonance for $e_2\ne 0$. But, such an extension is far beyond the scope of this paper, since we need additional inputs, {\it e.g.}
a detailed modeling of the outer precession rate ${\dot \varpi}_2$. We remain this issue as a future work.

\subsection{Averaged Hamiltonian}

To  geometrically explain the interesting behaviors in Fig. 1, we  briefly describe the employed Hamiltonian.
% temporarily ignoring radiation reaction. 
For studying the long-term evolution of the triples, we take the double averages of the original three-body  Hamiltonian with the inner and outer mean anomalies %following the von Zeipel transformation 
\cite{gs,Ford et al.(2000)}. After appropriate scaling, we obtain the relevant Hamiltonian    \cite{Blaes et al.(2002),Naoz et al.(2013)}
\beq
\mathcal{H}_{\rm T, 1PN}=\mathcal{H}_{\rm qp}+\mathcal{H}_{\rm 1PN}
\eeq
  with 
\beqa
\mathcal{H}_{\rm qp}&\equiv&-3G_{1}^{2}-15\frac{J_{1}^{2}}{G_{1}^{2}}-15(1-G_{1}^{2})\left(1-\frac{J_{1}^{2}}{G_{1}^{2}}\right)\cos2g_{1}, \nonumber\\
\\
\mathcal{H}_{\rm 1PN}&\equiv&-\frac{3\gamma}{G_{1}}. 
\eeqa
Here,  $\mathcal{H}_{\rm qp}$ is the Newtonian quadrupole term due to the outer tertiary, and  is obtained by perturbatively evaluating the tidal field of the tertiary $m_2$ with the expansion parameter $(a_1/a_2)\ll1$. The subsequent-order (octupole) term vanishes for $e_2=0$.  The 1PN term $\mathcal{H}_{\rm 1PN}$ leads to an apsidal precession of the inner orbit \cite{Fabrycky & Tremaine(2007)}. 
%We also put  $G_{1}\equiv \sqrt{1-e_{1}^{2}}$ that is defined in the range $J_{1}\leq G_{1}\leq1$.

The dynamical degree of freedom of this Hamiltonian is only the canonical variables $(g_{1}, G_{1})$. Below, as supported by the numerical experiments, we put $a_{1}=$ const and $J_{1}=$ const (ignoring the inner radiation reaction) \footnote{Also, our Hamiltonian $\mathcal{H}_{\rm T, 1PN}$ is independent of their conjugate variables.}. The decreasing  parameter $\gamma$ is defined as
\beq
\gamma\equiv\frac{16(m_{0}+m_{1})^{2}a_{2}^{3}}{(m_{2}a_{1}^{4})}\propto \frac{a_2^3}{a_1^4} \lmk \propto \frac{T_{\rm KL}}{T_{\rm 1PN}}\rmk,
\eeq
and contains the information of the outer orbits in $\mathcal{H}_{\rm T, 1PN}$. The 1PN effect gradually becomes weaker, relative to the  tidal field of the infalling tertiary.

In the $(g_{1}$, $G_{1})$ coordinate, the angle variable $g_{1}$ is singular at $G_{1}$ = $J_{1}$ and 1, since $g_{1}$ loses its geometrical meanings there (either $I=0$ or $e_{1}=0$). These coordinate singularities should be regarded as two degenerate points, and  can be appropriately handled with the canonical transformations \cite{Ivanov et al.(2005)}; 
\beqa
(x', y')=\sqrt{2(G_{1}-J_{1})}(\cos g_{1}, \sin g_{1})\\
(x, y)=\sqrt{2(1-G_{1})}(\cos g_{1}, \sin g_{1})
\eeqa
  As expected from geometric symmetry, these two points are fixed points in the phase space, irrespective of the parameter $\gamma$.

\subsection{Geometrical approach}

For our geometrical approach, we need to understand how the phase-space structure  %(such as fixed points and separatrixes)
 changes along with the slowly decreasing parameter $\gamma$. From general arguments on Hamiltonian systems, a fixed point must be either a center (stable fixed point) or saddle (unstable fixed point) \cite{strogatz}.

In Fig. 2, we show the phase-space structure of the Hamiltonian $\mathcal{H}_{\rm T, 1PN}(g_{1}, G_{1}; \gamma)$ for  $\gamma=14.89, 10.03$ and 7.37. First, we focus on the solid black curves (contour samples of $\mathcal{H}_{\rm T, 1PN}$), the black points (unstable fixed points), and the red broken curves (separatrixes).  These elements evolve in the following order.

\begin{figure*}[!htb]
\begin{center}
\includegraphics[width=12cm]{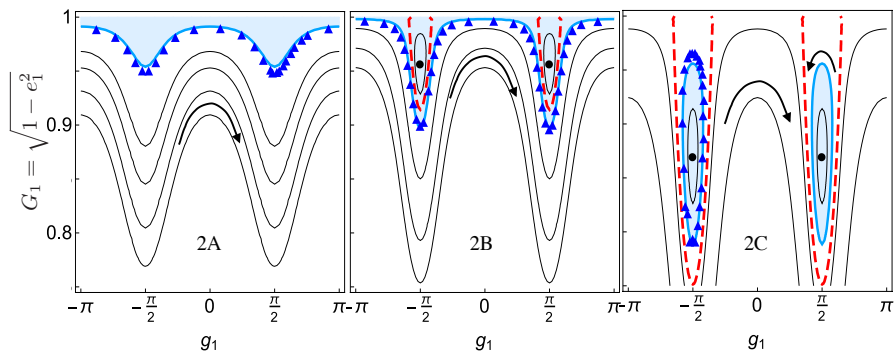}
\caption{
PN evolution of  the phase-space of the inner binary  with $J_1=0.150$ (plotted in the limited range $0.75\le G_1\le1$). Figs. 2A, 2B and 2C correspond to $\gamma$=14.88, 10.03 and 7.37 (decreased by the tertiary). The following  two types of information are presented simultaneously in each panel. [1. the overall phase-space structure of the Hamiltonian $\mathcal{H}_{\rm T,1PN}$] solid black curves: contour samples of $\mathcal{H}_{\rm T,1PN}$, black points: stable fixed points, red dashed curves: separatrixes,  and arrows: directions of flows. [2. evolution of the specific system s-I\hspace{-.1em}I\hspace{-.1em}I] The blue regions have equal area $S_{\rm I\hspace{-.1em}I\hspace{-.1em}I}=0.129$ (corresponding to the adiabatic invariant).  The triangles represent the results from the direct three-body calculation for s-I\hspace{-.1em}I\hspace{-.1em}I.}
\label{fig:phase}
\end{center}
\end{figure*}

(P1) For $\gamma>12-20J_{1}^{2}\equiv\gamma_{\rm c}$ (Fig. 2A), we only have circulating trajectories. Accordingly, the fixed points at $G_{1}=1$ and $J_{1}$ are both stable. The latter remains stable during whole.

(P2) At the incident  $\gamma=\gamma_{\rm c}$ (between Figs. 2A and 2B),   the fixed point $G_{1}=1$ transits from a stable point to an unstable point (pitchfork bifurcation \cite{strogatz}). For $J_1=0.150$,  this bifurcation is at $\gamma_{\rm c}=11.55$  (nicely corresponding to $a_{2}\simeq 5238$ in Fig. 1A).

(P3) For $\gamma<\gamma_{\rm c}$ (Figs. 2B and 2C), we simultaneously have circulating and librating trajectories, segregated by the separatrixes. As $\gamma$ decreases, the stable points at $g_{1}=\pm \pi/2$ take smaller $G_{1}$. The area inside the separatrixes increases monotonically.

Then,  how can we trace the evolution of individual systems in the sequence of Figs.2A-2C?  Here, we explain a powerful tool. Let us consider a periodic motion described by a Hamiltonian $\mathcal{H}(q, p; \lambda)$, including a parameter $\lambda$.   If the parameter $\lambda$ changes more slowly than the period of the motion, the area surrounded by the periodic trajectory $S\equiv\oint p dq$  is conserved and called an adiabatic invariant \cite{LL1,gs}. 
%Geometrically,  $S$ represents the area surrounded by the periodic trajectory in the phase space.

Now, let us trace the evolution of the system s-I\hspace{-.1em}I\hspace{-.1em}I in the phase-space, based on the adiabatic invariant.
As in Fig. 2A, this system initially circulates in the $(g_{1}, G_{1})$ plane, but it is periodic in the $(x, y)$ plane with the calculated  adiabatic invariant $S=0.129\equiv S_{\rm I\hspace{-.1em}I\hspace{-.1em}I}$. Here the area $S_{\rm I\hspace{-.1em}I\hspace{-.1em}I}$ corresponds to the  blue region in Fig. 2A. Because of the area conservation, at the stage of Fig. 2B, s-I\hspace{-.1em}I\hspace{-.1em}I is expected to stay circulating on the thick blue curve.

But, as $\gamma$ decreases, the total area inside the two separatrixes increases, and it eventually becomes identical to  $S_{\rm I\hspace{-.1em}I\hspace{-.1em}I}$ between Fig. 2B and 2C (around $\gamma\sim8.89$). At the separatrix crossing,  $e_{1}$ momently vanishes, and   the system s-I\hspace{-.1em}I\hspace{-.1em}I starts to librate either around $g_{1}=\pi/2$ or $-\pi/2$ (equal probability from the symmetry), due to the continuity of the trajectory.   Associated with this disjunction, the adiabatic invariant $S$ falls to one-half $S_{\rm I\hspace{-.1em}I\hspace{-.1em}I}/2$. But this gap is not contracting with the adiabatic invariance, since the period becomes infinite for the trajectory just on the separatrix. After crossing the separatrix, we can again follow the system, using the area conservation around one of the two stable points. For example, the averaged value of  $e_{1}$ increases with time. In this manner, we can geometrically  understand the evolution of s-I\hspace{-.1em}I\hspace{-.1em}I.

Until now, we have discussed the phase-space mapping of the specific system s-I\hspace{-.1em}I\hspace{-.1em}I.
Generally, in the mapping from Fig.2A to 2C, a system initially closer to $G_1=1$  is transported to a librating trajectory closer to one of the two stable fixed points (crossing the separatrix earlier).  As the present mapping is thus simple and symmetric, we can justify the area conservation also by applying  Liouville's theorem to certain group of systems ({\it e.g.} these composing the blue region in Fig. 2A).  Indeed, in Fig. 2, the total area of the blue region is the same, before and after the separatrix crossing.

Next, we examine the validity of our  approach, using the direct three-body calculations. In Fig. 2, the triangles represent the  numerical results for s-I\hspace{-.1em}I\hspace{-.1em}I. The associated area is conserved at high precision.  
Assuming the area conservation, we can inversely predict the evolution of $e_{1}$ easily.   In Fig. 1B, with the blue points, we represent the maximum and minimum values of $e_{1}$ at each $a_{2}$. They well reproduce the direct calculations \footnote{At the very end of the run, there exist  mismatches caused by   
breaks-down of the averaging method  for $e_1\sim 1$ \cite{Katz & Dong(2012),Bode & Wegg(2014),Seto:2013wwa}.}.

  In the case of s-I (with much smaller area $S_{\rm I}\ll S_{\rm I\hspace{-.1em}I\hspace{-.1em}I}$), the trajectory turns  into libration  soon after (P2)  and it stays close to the stable fixed point throughout Fig. 2B and 2C. This is why the distinctive  features of the pitchfork bifurcation (originally associated with the fixed points) appear in Fig. 1A.

\section{ Newtonian potential of   Stellar Clusters }

We next apply our approach to the evolution of individual stars in a stellar cluster around a central  MBH $m_0$ with an outer infalling tertiary MBH $m_2$.     We deal with component stars for which (i) the dominant force is the Newtonian attraction by the central MBH $m_0$, but (ii) its 1PN correction would be smaller than  the Newtonian potential of the cluster itself. Here, we ignore the 1PN effect for simplicity, but it could play important roles for stars with small pericenter distances (see {\it e.g.}  \cite{Merritt(2013)}). 
%(denoting the components by $(m_0$-$m_1)$-$m_2$ as before). 
In this section, we consider outer MBH that have larger separation (still on circular orbit) than  the previous  PN systems (see \S III.B for parameters in the physical units).
We do not specify the physical processes for the outer orbital decay, but the infall rate is assumed to be sufficiently small  ({\it e.g.} after the stall of dynamical friction \cite{Merritt(2013)}).    We also disregard dissipative effects for the inner orbit ({\it i.e.} setting $a_{1}={\rm const}$ and $J_{1}={\rm const}$).  The relaxation processes would be discussed in \S III.B (see {\it e.g.} \cite{Alexander:2005jz,Li et al.(2015),RT}).

\subsection{Geometrical approach}
We first present the averaged Hamiltonian. The  cluster is  initially assumed to be  spherical with the local density profile $\rho(r)=\rho_{1}(r/a_{1})^{-\beta}\,(\beta<3)$ around the distance $r=a_{1}$ in interest ($\rho_1$: the mean density there). Ignoring the back-reaction (time variation) to the stellar potential,  the scaled Hamiltonian is written as 
\beq
\mathcal{H}_{\rm T, SP}=\mathcal{H}_{\rm qp}+\mathcal{H}_{\rm SP}.
\eeq
 Here, $\mathcal{H}_{\rm SP}$ is the term due to the stellar potential
\begin{eqnarray}
\mathcal{H}_{\rm SP}=\eta e_{1}^{2}\left[1+\beta(-1+\beta)\left(\frac{e_{1}^{2}}{16}+\mathcal{O}(e_{1}^{4})\right)\right],
\end{eqnarray}
which is obtained by perturbatively expanding a hypergeometric function with $e_{1}=\sqrt{1-G_{1}^{2}}$ \cite{Merritt(2013)}. 
In our geometric approach, we neglect the small higher-order correction $o(e_1^2)$ that vanishes identically at $\beta=0$ and 1 (though exactly handled  later with $\beta=3/2$).
Namely, we put $\mathcal{H}_{\rm SP}=\eta (1-G_1^2)$ in terms of the canonical variable $G_1$.
 Then, all the information of the cluster is included in the 
single decreasing parameter 
\beqa
\eta\equiv\frac{16\pi\rho_{1}a_{2}^{3}}{m_{2}},
\eeqa
 with $ \dot{a}_{2}\equiv \p_t a_2<0$. 

The precession timescale due to the stellar potential $T_{\rm SP}$ \cite{Merritt(2013)} is
\beqa
T_{\rm SP}= \frac{(3-\beta)m_0}{4\pi \rho_1 a_{1}^{3}}P_1. \label{sp}
\eeqa
Then, we can represent $\eta$ by using $T_{\rm KL}$ and $T_{\rm SP}$ as
\beqa
\eta=\frac{12\pi}{3-\beta}\frac{T_{\rm KL}}{T_{\rm SP}}.
\eeqa

\begin{figure*}[!htb]
\begin{center}
\includegraphics[width=16cm]{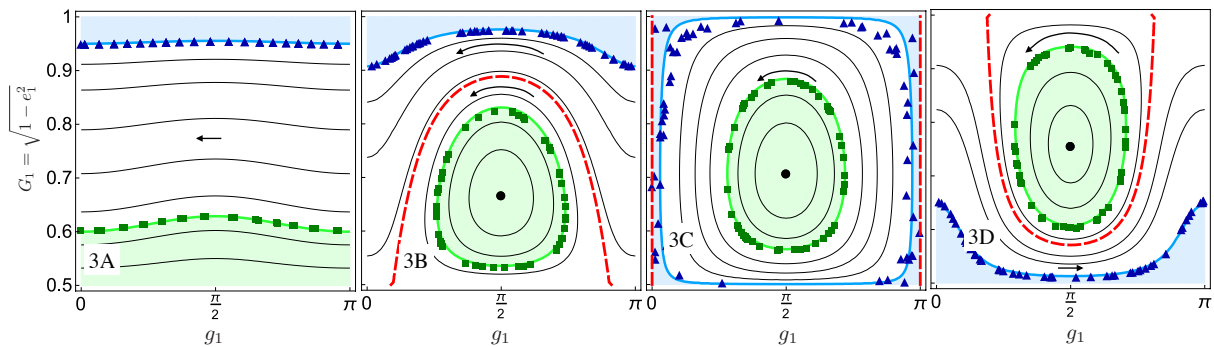}
\caption{
Evolution of the phase-space for stars in a stellar cluster around a MBH (fixed at $J_{1}=0.50$). Figures 3A, 3B, 3C and 3D are for $\eta$=200, 20, 12 and 5 decreased by an infalling tertiary MBH. Because of the symmetry, we only plot the range  $0\le g_{1}\le\pi$.  As in Fig. 2, the following two types of information are simultaneously plotted. [1. the phase-space structure of  $\mathcal{H}_{\rm T, SP}$] The symbols (solid black curves, black dots, red dashed curves and arrows) are defined in the same manner as  Fig. 2. [2. evolution of the two representative stars s-I\hspace{-.1em}V (blue) and s-V (green)]. The areas of the colored regions are respectively the same. The triangles and squares are obtained by numerically integrating the Hamilton equations with  $\beta=3/2$ and $\dot{\eta}$=$-10^{-3}$.}
\label{fig:ps2}
\end{center}
\end{figure*}

Next, we explain the evolution of the phase-space structure, following Fig. 3 (see also \cite{Ivanov et al.(2005)}).  The most remarkable point below is the bifurcation (P4') at $\eta_{\rm c2}\equiv 12$ (independent of $J_1$!).

(P1') For $\eta>-18+30/J_{1}^{2}\equiv\eta_{\rm c1}$ (Fig. 3A), we only have circulating trajectories in the retrograde direction.

(P2') At $\eta=\eta_{\rm c1}$ (between Figs. 3A and 3B), we have a pitchfork bifurcation similar to (P2). The fixed point $G_{1}=J_{1}$ turns  from a stable point to an unstable point ( in contrast to $G_1=1$ for (P2), see also Appendix A).  The new stable point (black point in Fig. 3B) is generated at $g_{1}=\pi/2$.

(P3') For $\eta_{\rm c2}\equiv12<\eta<\eta_{\rm c1}$ (Fig. 3B), we have librating trajectories inside the separatrix (red dashed curve). The inner area  increases with time. 

(P4') At $\eta=\eta_{\rm c2}$ (Fig. 3C),  due to a cancellation between $\mathcal{H}_{\rm qp}$ and $\mathcal{H}_{\rm SP}$, the Hamiltonian  $\mathcal{H}_{\rm T, SP}$ becomes independent 
of $G_1$ on the two vertical lines $g_1=0$ and $\pi$. This is a quite peculiar state in which fixed points spread fully on the two lines. At this moment, the stability of the two points $G_1= J_1$ and 1 interchange, and only librating trajectories exist.

(P5') For $\eta_{\rm c3}\equiv-18+30J_{1}^{2}<\eta<\eta_{\rm c2}$ (Fig. 3D), we again have both librating and circulating trajectories. The latter now have prograde motion. The area inside the separatrix decreases.

For $J_{1}>\sqrt{3/5}$, we have $\eta_{\rm c3}>0$. Thus, if $\eta$ decreases down to $\eta_{\rm c3}$, the stable point at $g_1=\pi/2$ disappears, and the fixed point $G_1=1$ turns to be stable (inverse of the bifurcation at (P2)). We then have only circulating trajectories.

Now, using the area conservation,  we follow the evolution of two representative  stars s-I\hspace{-.1em}V and s-V in the time sequence of Fig. 3. Their initial conditions ($g_{1}$, $G_{1}$) at $\eta=200$ are (0,0.95) for s-I\hspace{-.1em}V, and (0,0.6) for s-V both with $J_1=0.5$.  In Fig. 3A, we can effectively regard the areas of the colored regions (blue: s-I\hspace{-.1em}V and green: s-V) as  conserved quantities.  Conserving these areas, the two systems would evolve as shown with the thick blue/green curves in Figs. 3B-3D.  Note that, at the first separatrix crossing during (P3'), the maximum eccentricity ({\it i.e.} $G_1=J_1$) is surely realized (in contrast to $G_1=1$ for the 1PN case).

Interestingly, the magnitude of their eccentricities is inverted between Fig. 3A and 3D.  This is because of the passage of the peculiar state (P4'). Consequently,  nearly circular orbits $(G_1\sim 1)$ would be quickly converted   to eccentric ones ($G_1\sim J_1$) around $\eta=12$. This sharp transition is quite different from the more gradual evolution in the PN case, and would be less likely to be prevented by other processes.

If the cluster initially has an isotropic velocity distribution, the stars accordingly have a homogeneous density distribution in the $(g_1,G_1)$ plane \cite{GD}. Then, because of Liouville's theorem, the homogeneity is unchanged in the subsequent stages in Fig. 3, and our assumption about the stationarity of the cluster potential would be a good approximation.  Meanwhile, for an initial distribution biased to small eccentricities, the secure realization of the maximum eccentricity during (P3') could help to refill the loss cone, and might be very important astrophysically.

In Fig. 3, we additionally present the results obtained by numerically solving the Hamilton equations;  
\beq
\dot{G}_{1}=-\frac{\partial \mathcal{H}_{\rm T,SP}}{\partial g_{1}},~~ \dot{g}_{1}=\frac{\partial \mathcal{H}_{\rm T, SP}}{\partial G_{1}}
\eeq
 with $\beta=3/2$ and $\dot{\eta}=-10^{-3}$ (including $o(e_1^2)$-term for $\mathcal{H}_{\rm  SP}$). Except for some fluctuations around $\eta=12$,  our simple geometrical predictions fit the numerical results well. 

\subsection{Effects of Relaxation}
%Until now, we assume that the stellar potential is smooth. However the distribution of stars in the stellar cluster is discrete, and the diffusion of energy and angular momenta (due to two-body scattering, resonant relaxation and so on) occurs. These effects are crucial for the geometrical approach ({\it i.e.} adiabatic invariant) if the infall time of the outer MBH is longer than those characteristic timescales. In this section, by estimating those characteristic timescales, we discuss the situation in which these effects could be ignored.
Until now, we have neglected the diffusion of the energy and angular momentum associated with the discreteness of the stellar distribution. In this subsection, we estimate the characteristic timescales of the relevant relaxation processes. 

\subsubsection{Model parameters}
%In the following, we use the physical unit and set the fiducial values of parameters. We choose the masses at $m_0=10^8 M_{\odot}$, $m_1=M_{\odot}$ and $m_{2}=q m_{0}$ ($q<1$). We assume the mass of the stellar cluster inside $r$ is $M_{\ast}(r)=2m_0 (r/r_{\rm h})^{3/2}$. The influenced radius $r_h$ is $r_{\rm h}=Gm_{0}/\sigma^2=14.0{\rm pc}(m_0/10^8 M_{\odot})^{0.645}$, where we eliminate the 1 dimensional velocity dispersion $\sigma$ by using $M_{\bullet}$ - $\sigma$ relation $M_{\bullet}=2.09\times10^8 M_{\odot}(\sigma/200{\rm km/s})^{5.64}$ \cite{MM}. 
%Then, the density profile is $\rho(r)=3m_0/(4\pi r_{\rm h}^{3})(r/r_{\rm h})^{-3/2}$ (corresponding to set $\beta=3/2$).
To begin with, we  set the fiducial model for  the overall cluster density profile as
\beqa
\rho(r)=\frac{3m_0}{4\pi r_{\rm h}^{3}}\left(\frac{r}{r_{\rm h}}\right)^{-3/2},\label{3/2}
\eeqa
with the slope $\beta=3/2$ \cite{OL(2009)}. Here $r_{\rm h}$ is the influenced radius $r_{\rm h}=Gm_{0}/\sigma^2$ with  the one-dimensional velocity dispersion $\sigma$ that is assumed to satisfy the $M_{\bullet}$ - $\sigma$ relation \cite{MM};
\beq
M_{\bullet} (=m_0) =2.09\times10^8\lmk \frac\sigma{200{\rm km/s}}\rmk^{5.64}M_{\odot}.
\eeq

From Eqs. (\ref{sp}) and (\ref{3/2}), the timescale $T_{\rm SP}$ of the mass precession is given by 
\beqa
T_{\rm SP}= 2.58\times10^5\left(\frac{m_{0}}{10^{8}M_{\odot}}\right)^{0.47}{\rm yr}
\eeqa
without depending on $a_1$ (for $\beta=3/2$). 

In the standard picture of MBH binary evolution  \cite{Merritt(2013)}, the orbital decay timescale $T_{\rm dec}\equiv a_2/{\dot a}_2$ is considered to increase sharply around the hard-binary separation 
\beqa
a_{\rm h}&\equiv&\frac{m_{2}}{m_0+m_2}\frac{r_{\rm h}}{4}\nonumber\\
&=&3.50\frac{q}{q+1}\left(\frac{m_0}{10^8M_{\odot}}\right)^{0.645}{\rm pc}
\eeqa
with $q\equiv m_2/m_0<1$.
The subsequent orbital decay would strongly depend on the exterior of the MBH binary ({\it e.g.} triaxial potential \cite{Khan et al.(2011),Vasiliev et al.(2014)}), and the actual timescale $T_{\rm dec}$ is still under active discussion. We are specifically interested in the orbits of stars around the peculiar bifurcation  at $\eta=12$. For applying  the adiabatic invariance, the decay timescale $T_{\rm dec}$ must be smaller than  the relaxation timescales, but larger than $T_{\rm KL}$ that is explicitly given as
\beqa
T_{\rm KL}&=&\frac{5.45\times10^5}{q}\left(\frac{m_0}{10^8 M_{\odot}}\right)^{-1/2}\nonumber\\
& &\times\left(\frac{a_1}{0.3{\rm pc}}\right)^{-3/2}\left(\frac{a_2}{3.5{\rm pc}}\right)^3 {\rm yr}.
\eeqa

\subsubsection{Resonant Relaxations}

Now, we discuss the timescales for the relaxation processes. As is well-known for nuclear star clusters, the timescale for the two-body relaxation is generally much larger than that for the resonant relaxation. Therefore, in the following,    we concentrate on the resonant relaxation.

  We firstly summarize the basic aspects of the  resonant relaxation for a conventional system composed only by a central MBH and the associated stellar clusters   \cite{Merritt(2013)}.  Then, we discuss how the resonant relaxation would be modified by the outer MBH $m_2$.

In a spherical  stellar cluster around a central MBH $m_0$, the individual stars maintain their Keplerian orbits for the precession timescale $T_{\rm SP}$. The scalar resonant relaxation is the long-term accumulation of the gravitational interaction between these  Keplerian orbits of the short-coherence time $T_{\rm SP}$.  Both the magnitude and orientation of the angular momenta of the stars change with the characteristic timescale  \cite{Merritt(2013)}
\beqa
T_{\rm rr,s}&=&\frac{m_{0}}{m_{1}}P_{1}\nonumber\\
&=&1.6\times10^{9}\left(\frac{1M_{\odot}}{m_1}\right)\left(\frac{m_{0}}{10^{8}M_{\odot}}\right)^{1/2}\left(\frac{a_1}{0.3 {\rm pc}}\right)^{3/2}{\rm yr}. \nonumber\\
\label{rrs}
\eeqa

Meanwhile, the vector resonant relaxation  should be regarded as  interaction between the mass annulus obtained after the orbital averaging over the precession timescale $T_{\rm SP}$. It changes only the direction of the angular momentum (due to the symmetry) with the characteristic time  \cite{Merritt(2013)}
\beq
T_{\rm rr,v}\sim \frac{T_{\rm rr,s}}{\sqrt{N}},
\eeq 
where $N$ is the number of stars around the semimajor axis in interest.

Next, we discuss the impacts of the additional outer MBH $m_2$ on the two-types of resonant relaxations.   Now, the individual Keplerian  orbits maintain their shapes for  ${\rm min}[T_{\rm KL}, T_{\rm SP}]$. Therefore, for $T_{\rm KL}\lsim T_{\rm SP}$, the timescale of the scalar resonant relaxation would be a factor of $T_{\rm SP}/T_{\rm KL}=\lmk \eta/12\rmk^{-1} \pi/\lmk3-\beta\rmk$ times larger than $T_{\rm rr,s}$ in Eq.(\ref{rrs}). This factor is of order unity for $\eta\sim 12$,

Here, it is important to note that the outer MBH also generates precession of the longitude of the ascending node $\Omega_1$ of the stars with the characteristic timescale $T_{\rm KL}$ (because $|{\dot \Omega}_1|=|\partial \mathcal{H}_{\rm T,SP}/\partial J_{1}|\sim |\partial \mathcal{H}_{\rm T,SP}/\partial G_{1}|=|{\dot g}_1|$).  Therefore, around $\eta\sim12$, the vector resonant relaxation should be regarded as gravitational interaction between axisymmetric (torus-like)  mass distributions with the common symmetric axis (rather than the inclined annulus for the conventional case). But there is no transfer of torque between them (due to the symmetry), and, consequently,  the vector resonant relaxation becomes ineffective around the regime $\eta\sim12$.

We conclude that, even though the decay timescale $T_{\rm dec}=a_2/{\dot a}_2$ is uncertain at $a_2<a_{\rm h}$, it should be smaller than  $T_{\rm rr,s}$ in Eq.(\ref{rrs}) for applying the adiabatic invariance to the bifurcation around $\eta=12$.

\section{Discussions}

We have developed a  geometrical approach for the adiabatic evolution of hierarchical triple systems including  MBH binaries on circular orbits.  We analyzed the basic profile ({\it e.g.} fixed points and separatrixes ) of the phase-space structure of the corresponding Hamiltonian.  In addition, we apply the arguments of the adiabatic invariance to follow the time evolution of individual systems.  

At first, we demonstrated the validity of our approach using a simple relativistic systems that can be also examined by direct numerical calculations. Then, we analyzed more realistic systems, star clusters associated with inspiraling MBH binaries.
We found that the  eccentricities of the stars could evolve quite interestingly. This is caused by the peculiar bifurcation patterns induced by the Newtonian potential of stellar clusters, in sharp contrast to the precedent relativistic systems.  Our simple geometrical approach enables us to clearly understand how the inner orbits are deformed,  in response to the time variation of the related phase-space structure. It would be applicable to the KL mechanism in various astrophysical contexts, possibly including viscous effects to the inner orbit at least for a short time.

\begin{acknowledgments}
We would like to thank M. Iwasawa,  H. Nakano and A. Tanikawa for helpful conversations. This work was supported by JSPS (24540269, 15K05075) and MEXT (24103006).
\end{acknowledgments}

\appendix

\section{Generation of a New Fixed Point}
As shown in Figs. 2 and 3, the phase-space evolution is largely different between the two Hamiltonians 
${\cal H}_{\rm T,1PN}$ and ${\cal H}_{\rm T,SP}$ given in Eqs. (6) and (12). 
When relatively increasing the quadrupole term ({\it i.e.} $\gamma,\eta=+\infty \to 0$), 
one of the remarkable contrasts is the emergence of the new fixed point either at $G_1=1$
 (the epoch (P2) for 1PN) or at $G_1=J_1$ (the epoch (P2\rq{}) for SP) both with $g_1=\pm \pi/2$. In this appendix, we discuss how
 this contrast is related to the sign of the apsidal precession (1PN: prograde and SP: retrograde).

In the present case, the condition for the fixed point is formally expressed as 
\beqa
\frac{\partial \mathcal{H}_{\rm T}}{\partial G_1}\Bigr|_{g_1=\pi/2}=0
\eeqa
for ${\cal H}_{\rm T}={\cal H}_{\rm T,1PN}$ and ${\cal H}_{\rm T,SP}$. This equation is physically equivalent to
 ${\dot g}_1=0$ for the apsidal precession rate,  representing the balance 
 between the quadrupole effect and the competing 1PN or SP effect.
In concrete terms, we collectively have
\beqa
Q(G_1)\equiv -36G_1+60\frac{J_{1^2}}{G_{1}^3}=-kG_{1}^{-\alpha}.
\label{fp}
\eeqa
Here, the function $Q(G_1)$ originates from the quadrupole term and monotonically decreases in the relevant range
 $J_1\le G_1\le 1$. The right hand side shows the apsidal precession rates induced by the 1PN ($\alpha=2$ and $k(=3\gamma)>0$) 
and the SP 
($\alpha=-1$ and $k(=-2\eta)<0$).

Now, for simplicity, we assume $J_1<\sqrt{3/5}$ (identical to the condition for the standard KL-libration).
Then, we have $Q(G_1=J_1)>0$ and $Q(G_1=1)<0$. Furthermore,  with respect to the two types of 
variations  $k=+\infty \to 0$ or $k=-\infty \to 0$,   either of two solutions; $G_1=J_1$ or $G_1=1$ initially appears in Eq.(\ref{fp}), 
irrespective of the index  $\alpha \in [-1,2]$. 
Indeed,  depending only on the sign of 
$k$, we  have   $G_1=1$ (for $k=+\infty \to 0$) and $G_1=J_1$ (for $k=-\infty \to 0$).

This result explains the sharp contrast between (P2) and (P2\rq{}). Here, the sign of the apsidal precession is crucially important. 
Note also that, even if a small correction term is added to the right hand side of Eq.(\ref{fp}), this robust result is unchanged.

\end{document}